\documentclass[reprint,pra,aps,notitlepage,nofootinbib]{revtex4-2}
\usepackage{amsmath}
\usepackage{amssymb}
\usepackage{bm,braket}
\usepackage{float}

\usepackage{hyperref}
\usepackage{multirow}
\hypersetup{
     colorlinks   = true,
     citecolor    = blue
}
\def\be{\begin{equation}}
\def\ee{\end{equation}}
\def\bea{\begin{eqnarray}}
\def\eea{\end{eqnarray}}
\def\bes{\begin{eqnarray}}
\def\ees{\end{eqnarray}}
\def\bi{\begin{itemize}}
	\def\ei{\end{itemize}} 
	
\usepackage{amsthm}

\theoremstyle{definition}

\usepackage{tikz}
\usetikzlibrary{braids}

\begin{document}
\title{Combinatorial optimization with quantum imaginary time evolution }
	\author{Nora M. Bauer}
	\email{nbauer1@vols.utk.edu}
\affiliation{Department of Physics and Astronomy, The University of Tennessee, Knoxville, TN 37996-1200, USA}
\author{Rizwanul Alam}
\email{ralam4@vols.utk.edu}
\affiliation{Department of Physics and Astronomy, The University of Tennessee, Knoxville, TN 37996-1200, USA}
\author{James Ostrowski}
\email{jostrows@tennessee.edu}
\affiliation{
	Department of Industrial and Systems Engineering, The University of Tennessee, Knoxville, Tennessee  37996-2315, USA}
\author{George Siopsis}
	\email{siopsis@tennessee.edu}
\affiliation{Department of Physics and Astronomy, The University of Tennessee, Knoxville, TN 37996-1200, USA}
	 
\date{\today}
\begin{abstract}
We use Quantum Imaginary Time Evolution (QITE) to solve polynomial unconstrained binary optimization (PUBO) problems. We show that a linear Ansatz yields good results for a wide range of PUBO problems, often outperforming standard classical methods, such as the Goemans-Williamson (GW) algorithm. We obtain numerical results for the Low Autocorrelation Binary Sequences (LABS) and weighted MaxCut combinatorial optimization problems, thus extending an earlier demonstration of successful application of QITE on MaxCut for unweighted graphs.  We find the performance of QITE on the LABS problem with a separable Ansatz  comparable with $p=10$ QAOA, and do not see a significant advantage with an entangling Ansatz. On weighted MaxCut, QITE with a separable Ansatz  often outperforms the GW algorithm on graphs up to 150 vertices. 
\end{abstract}
\maketitle 
\section{Introduction}  

Many important combinatorial optimization problems can be mapped onto an Ising-type Hamiltonian. Solving a generic Ising model is an NP-hard problem \cite{BarahonaF1982Otcc}. It would be interesting to witness quantum algorithms outperform their classical counterparts in such problems, but demonstrating their superiority with NISQ devices has proved to be challenging. One of the quantum algorithms extensively studied on NISQ hardware is the Quantum Approximate Optimization Algorithm (QAOA) \cite{farhi2014quantum}. Another interesting approach was recently proposed that performs better on NISQ devices based on a quantum-enhanced classical optimization algorithm \cite{DupontQuantum-enhanced}. 

After mapping onto a Hamiltonian, the combinatorial optimization problem reduces to finding the ground state energy. Various methods have been employed to compute the ground state of a system, such as adiabatic evolution, quantum annealing and quantum imaginary-time evolution (QITE). QITE has been widely employed in the analysis of quantum many-body systems, serving as a valuable technique for diverse purposes, including computation of energy levels of multi-particle systems and generation of states at finite temperatures \cite{McArdle2019, beach2019making}. Given that evolution in imaginary time effectively reduces the system to zero temperature \cite{love2020cooling}, the ground state can be precisely prepared using QITE without the need for variational optimization. Nevertheless, in practice approximations are necessary due to limited computational resources, prompting the need for an approach involving variational calculus.

The progression in imaginary time $\tau$ is executed through the non-unitary operator $U(\tau) = e^{-\tau \bm{\mathcal{H}}}$, where $\bm{\mathcal{H}}$ denotes the Hamiltonian of the system of interest. Starting with an initial state featuring a non-zero overlap with the system's ground state, the evolved state converges towards the ground state as $\tau$ approaches infinity. Access to excited states is also attainable by selecting an appropriate initial state, one that is orthogonal to the ground state. Simulating QITE on a quantum computer is not straightforward, because $U(\tau)$ is a non-unitary operator. Motta \textit{et al.}\ \cite{motta2020determining} proposed a QITE algorithm that dispensed with the need for classical optimization, distinguishing it from QAOA \cite{farhi2014quantum}. Furthermore, it exhibited an advantage over a variational quantum eigensolver (VQE) by not employing ancilla qubits. The approach found practical application in the quantum computation of chemical energy levels on NISQ hardware \cite{gomes2020efficient, yeter2020practical, yeter2021benchmarking,2020arXiv201108137B} and in the simulation of open quantum systems \cite{kamakari2021digital}. The impact of noise on QITE in NISQ hardware was addressed in \cite{ville2021leveraging} using error mitigation and randomized compiling. Error mitigation was also addressed with a different method based on deep reinforcement learning \cite{cao2021quantum}. The reduction of quantum circuit depth for QITE through a nonlocal approximation was discussed in \cite{nishi2021}. The implementation of real and imaginary time evolution using compressed quantum circuits on NISQ hardware was demonstrated in \cite{PRXQuantum.2.010342}.

The QITE algorithm initiates by representing the Hamiltonian in local terms and employs Trotterization to implement $U(\tau)$. Subsequently, the non-unitary evolution for a short imaginary-time interval is approximately implemented with a unitary operator. This unitary operator is expressed in terms of Pauli spin operators, with coefficients determined from measurements on quantum hardware. This approach was employed in Ref.\ \cite{Alam2023} to solve combinatorial optimization problems. For the approximate unitary operator, an Ansatz was used that was linear in the Pauli operators and therefore its implementation required no entanglement of qubits. The method was applied to the MaxCut problem on thousands of randomly selected unweighted graphs with up to fifty vertices. Results compared favorably with the performance of classical algorithms, such as the greedy \cite{10.1007/3-540-63774-5_137,articlegreedy,10.5555/1347082.1347102,7133122} and Goemans–Williamson (GW) \cite{goemans1994879} algorithms. The overlap of the final state of the QITE algorithm with the ground state was also discussed as a performance metric, which is a quantum feature not shared by other classical algorithms. These results indicate that the linear QITE method is efficient and quantum advantage due to entanglement is likely to be found only at larger graphs ($N\gtrsim 100$) requiring deep quantum circuits which cannot currently be handled by NISQ hardware.

Given the demonstrated success of QITE based on a linear Ansatz  for MaxCut problems on unweighted  \cite{Alam2023}, it is crucial to assess its efficacy in more general polynomial unconstrained binary
optimization (PUBO) problems, which is a more general class of problems containing the more popular quadratic unconstrained binary optimization (QUBO) problems. This analysis is important for assessing the utility of NISQ devices.

Here we extend the results of Ref.\ \cite{Alam2023} to solve PUBO problems using QITE. We concentrate on two problems: (a) the MaxCut problem on weighted graphs, and (b) the Low Autocorrelation Binary Sequences (LABS) problem. In both cases we map the problem onto an Ising-type Hamiltonian. 

In the MaxCut problem on weighted graphs, we tested QITE with a separable linear Ansatz on graphs up to 150 vertices (qubits), and found that QITE often outperformed the GW algorithm, attaining an average AR $0.95$ for $N>100$ vertices. Even as the energy gaps between the ground and first excited state became small, QITE was able to find a state with an equal or lower energy than GW. It should be noted that QITE with a separable Ansatz can be simulated efficiently classically. Our results indicate that quantum advantage in combinatorial optimization problems will be hard to witness on NISQ devices.

In the LABS problem, complexity grows quickly for large $N$. Optimal solutions are only known for $N\leq66$, so LABS is a promising candidate for quantum advantage as no classical solutions currently exist \cite{shaydulin2023evidence}. The relevant regime where classical heuristics produce poor solutions is $N\approx 200$, so the number of qubits required for a solution to a classically intractable problem is on the order of hundreds of qubits.  In Ref.\  \cite{shaydulin2023evidence},  they obtained a scaling advantage over classical algorithms for problem size up to $40$ qubits using QAOA up to level $p=40$, obtaining exact solutions. We applied linear Ansatz QITE and obtained probability of measuring the ground state, P(GS), comparable with level $p=10$ QAOA results at relatively low circuit depth and hardware requirements. We also investigated the performance of an entangling quadratic Ansatz with QITE, but found no significant advantage over the linear Ansatz QITE for $N<10$.

Our discussion is organized as follows. In Section \ref{sec:2}, we discuss the QITE procedure and its hardware requirements. In Section \ref{sec:3}, we define the weighted MaxCut problem and give results on graphs up to 150 vertices (qubits). In Section \ref{sec:4}, we define the LABS problem and discuss how we solve it with QITE, and present results for problem size up to 28 qubits. Finally, in Section \ref{sec:V}, we summarize our results and discuss further research directions.  

\section{Method}  \label{sec:2}
In this Section, we review the QITE procedure introduced in Ref.\ \cite{Alam2023}, and discuss the hardware requirements of the algorithm. 

To implement QITE, we perform evolution in small imaginary time intervals $\tau$. 
In the zero temperature limit, the ground state of the Hamiltonian $\mathcal{H}$ for any state $\ket{\Psi}$ is given by 
\be \ket{\Omega} = \lim_{\beta\to\infty} \frac{e^{-\beta \bm{\mathcal{H}}} \ket{\Psi}}{\|  e^{-\beta \bm{\mathcal{H}}} \ket{\Psi} \|} \label{eq:QITE}\ee 
as long as $\braket{ \Omega |\Psi} \ne 0$. The Hamiltonian is defined on a graph $G \equiv (V,E)$ where $V$ ($E$) is the set of vertices (edges). The system consists of qubits lying on the vertices of the graph $G$.

We initialize the system in the state $\ket{\Psi[0]}$, which can be arbitrarily chosen, as long as it has finite overlap with the ground state of the system. Suppose that after $s-1$ steps we arrive at the state $\ket{\Psi[s-1]}$. In the $s$th step, we approximate the evolution of $\ket{\Psi[s-1]}$ in (small) imaginary time $\tau$ by the action of the unitary $e^{-i\tau A[s]}$, where $A[s]$ is a Hermitian operator.  Thus, after $s$ steps, we arrive at the state
\be \ket{\Psi [s]} = e^{-i\tau A[s]} \ket{\Psi[s-1]} = \prod_{s'=1}^s e^{-i\tau A[s']} \ket{\Psi[0]} \label{eq:approx}\ee
For the unitary updates, we adopt the \emph{linear Ansatz}
\be A[s] = \sum_{j\in V} a_j[s] Y_j \label{eq:8}\ee 
This unitary update is optimal when the coefficients $a_j[s]$ obey the linear system of equations \cite{Alam2023}
\be \bm{S}\cdot \bm{a} = \bm{b} \ , \ \ S_{ij}[s] = \braket{ Y_i Y_j } \ , \ \ b_j[s] = -\frac{i}{2} \braket{ [ \bm{\mathcal{H}} , Y_j] } \ee
where all expectation values are evaluated with respect to the state $\ket{\Psi[s-1]}$ obtained in the previous step. 

We choose the initial state to be the tensor product of eigenstates of $X$ and $Z$,
\be \ket{\Psi[0]} = \bigotimes_{j=1}^{|V|} \ket{s_j} \ , \ \ \ket{s_j} \in \{ \ket{0}, \ket{1}, \ket{+}, \ket{-} \}\ee 
where $\ket{\pm} = \frac{1}{\sqrt{2}} (\ket{0} \pm \ket{1})$, introducing no entanglement. Consequently, we have $\bm{S}=\mathbb{I}$, and therefore $\bm{a}=\bm{b}$. 
Since all unitary updates commute with each other, we may write the state after $s$ steps as
\be \ket{\Psi[s]} = e^{-i\tau \mathcal{A}[s]} |\Psi[0]\rangle \ , \ \ \mathcal{A}[s] = \sum_{s'=1}^s A[s'] \ee 
In terms of hardware requirements, at each QITE step we need only measure $\bm{b}$,
which requires $N$ measurements due to basis rotations. In the results shown in the following sections, the value of the small imaginary time parameter $\tau$ is chosen by increasing in steps of $\Delta\beta = \frac{\tau}{T}$, where $T\ll \tau$, starting from zero. At each increase, the cost function $\braket{\bm{\mathcal{H}}}$ is measured and the process is continued until the cost function (energy) starts increasing. At that point, the previous $\tau$ value is selected. We choose a maximum $\beta_{\mathrm{max}}$ and perform less than $\beta_{\mathrm{max}}/\Delta\beta$ measurements. 
Since the applications of $e^{-ia_j[s]Y_j}$ commute, the circuit depth does not scale with the number of steps.  The number of measurements scales linearly with the number of steps.  
  
\section{Weighted MaxCut} \label{sec:3}

Here we define the weighted MaxCut problem and the technique of imaginary-time-dependent (ITD) edges that we use to improve convergence. We present results of graphs up to 150 vertices (qubits), and find that QITE sometimes outperforms the classical GW algorithm \cite{goemans1994879}. 

Given a graph $G = (V, E)$ consisting of a set of vertices V and edges $E$ joining the vertices, the unweighted MaxCut problem on $G$ is the combinatorial optimization problem of partitioning $V$ into two disjoint sets such that the number of edges with endpoints in each set, $C$, is maximized ($C = C_{\mathrm{max}}$). It can be formulated as a Hamiltonian ground-state problem by associating a qubit with every vertex in $V$ and defining the Hamiltonian 
\be \bm{\mathcal{H}} =\sum_{(i,j)\in E}Z_iZ_j\ee 
where $Z_i$ is the Pauli $Z$ matrix acting on the $i$th qubit. The ground state energy $\mathcal{E}_0$ of $\bm{\mathcal{H}}$ is related to $C_{\mathrm{max}}$ by 
\be  C_{\mathrm{max}}=\frac{|E|-\mathcal{E}_0}{2}\ee 
For the weighted MaxCut problem, the edges $(i,j)\in E$ of the graph $G^w$ have associated weights $w_{ij}$, and the Hamiltonian is modified as 
\be \bm{\mathcal{H}}^w=\sum_{(i,j)\in E}w_{ij}Z_iZ_j\ee 
The ground state energy $\mathcal{E}_0^w$ is related to the maximum cut $C_{\mathrm{max}}^w$ as 
\be  C^w_{\mathrm{max}} = \frac{\sum_{(i,j)\in E} w_{ij}-\mathcal{E}_0^w}{2}\ee 
We compare the ground state energy obtained from QITE to the ground state energy obtained by  the classical GW algorithm \cite{goemans1994879}. Since for large graph sizes we cannot guarantee that GW produces the ground state energy, we define the approximation ratio (AR) as the energy obtained by QITE divided by the lowest energy produced by GW. 

As in \cite{Alam2023}, we use the method of ITD edges to improve convergence. This is done by interpolating between a Hamiltonian that corresponds to a subgraph of $G^w$ and $\bm{\mathcal{H}}$ using an ITD Hamiltonian $\mathcal{H}^w[s]$ given by: 
\be \bm{\mathcal{H}}^w[s]=\sum_{(i,j)\in E} w_{ij}[s]Z_iZ_j\ee 
A subgraph is selected for which the corresponding weights vanish initially, and then are increased with each step $s$ until $w_{ij}[s]\rightarrow w_{ij}$ for sufficiently many steps and $\mathcal{H}^w[s]\rightarrow\mathcal{H}^w$. 

We restrict our focus to $N$-vertex weighted Newman-Watts-Strogatz (NWS) graphs with $k=4$ and $p=0.5$, where the weights are selected from the uniform random distribution $(0,1]$. We apply the QITE algorithm with a linear Ansatz with 1 ITD edge, and measure the AR as the measured energy from QITE divided by the lowest energy obtained from GW. We also measure the probability of obtaining this lowest energy value as the probability of obtaining the ground state, P(GS). 

\begin{figure}[ht!] 
    \centering
    \includegraphics[scale=0.65]{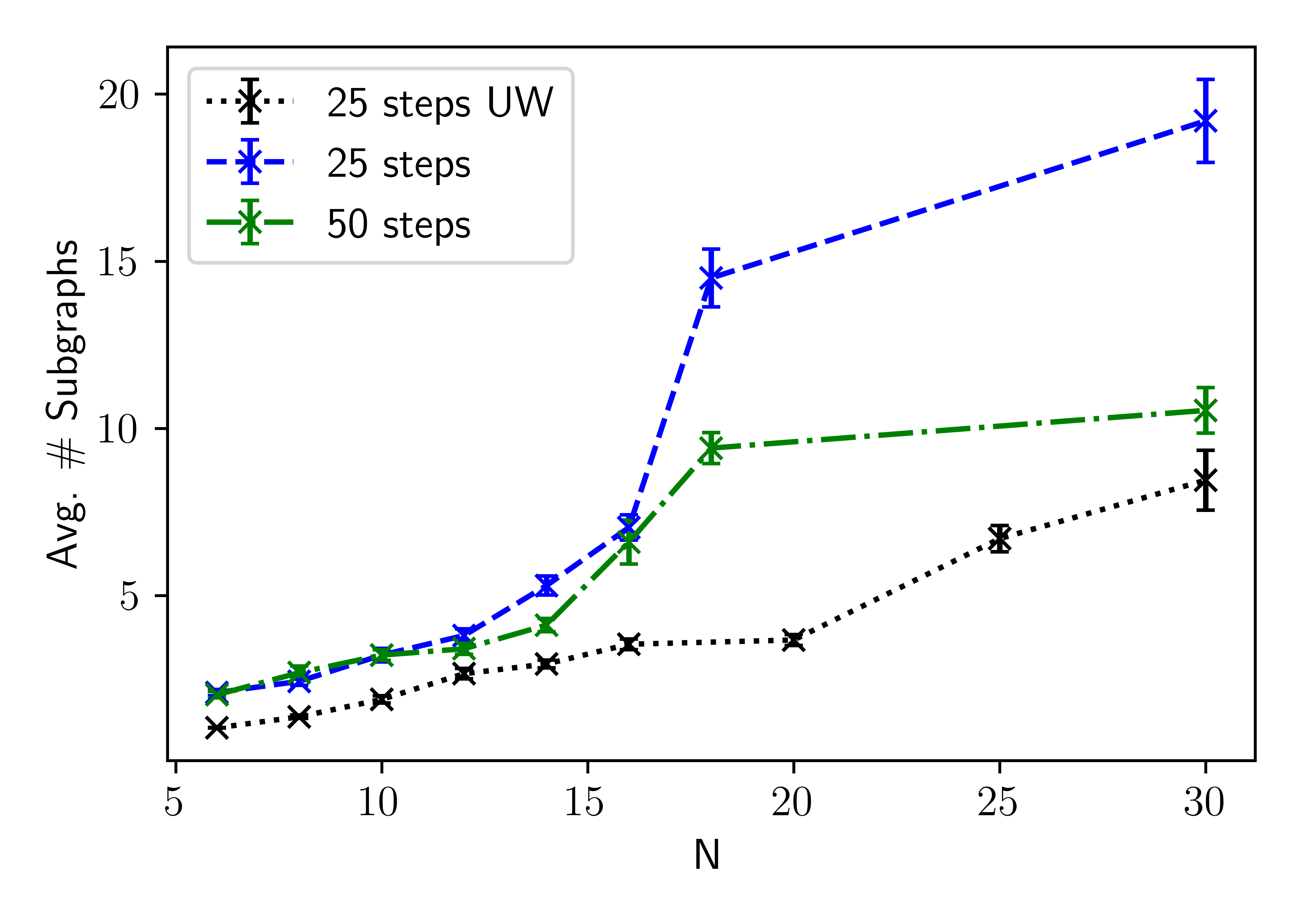} 
\caption{QITE results for the weighted MaxCut problem showing the average number of random subgraph choices required for $N$ qubit graphs to obtain P(GS)$>0.995$. Results are shown for 25 QITE steps on unweighted (black dotted line) and weighted (blue dashed line) graphs, as well as 50 QITE steps on weighted graphs (green dashed-dotted line). The bars denote 1 standard error over 250 trials.  }   
    \label{fig:f1A} 
\end{figure}  
\begin{figure}[ht!] 
    \centering
    \includegraphics[scale=0.65]{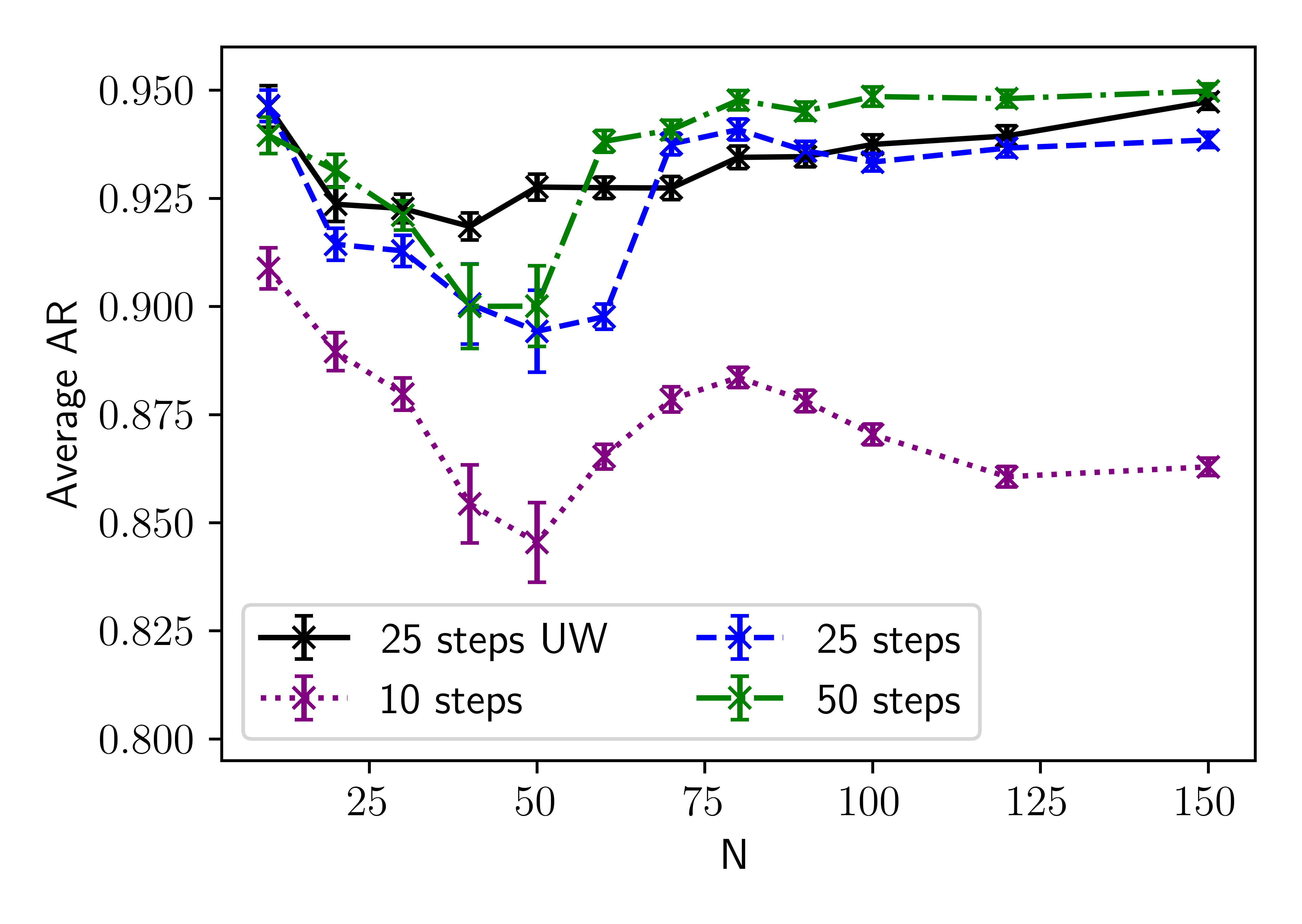} 
\caption{QITE results for the weighted MaxCut problem showing the average approximation ratio, computed by dividing the energy from QITE by the energy from GW, for single subgraph trials for $N$-qubit weighted graphs. Results are shown for 10 (dotted purple), 25 (dashed blue), and 50 (dashed-dotted green) QITE steps, in addition to 25 steps for the unweighted (UW) case (solid black). }   
    \label{fig:f1B} 
\end{figure}  
\begin{figure}[ht!] 
    \centering
    \includegraphics[scale=0.65]{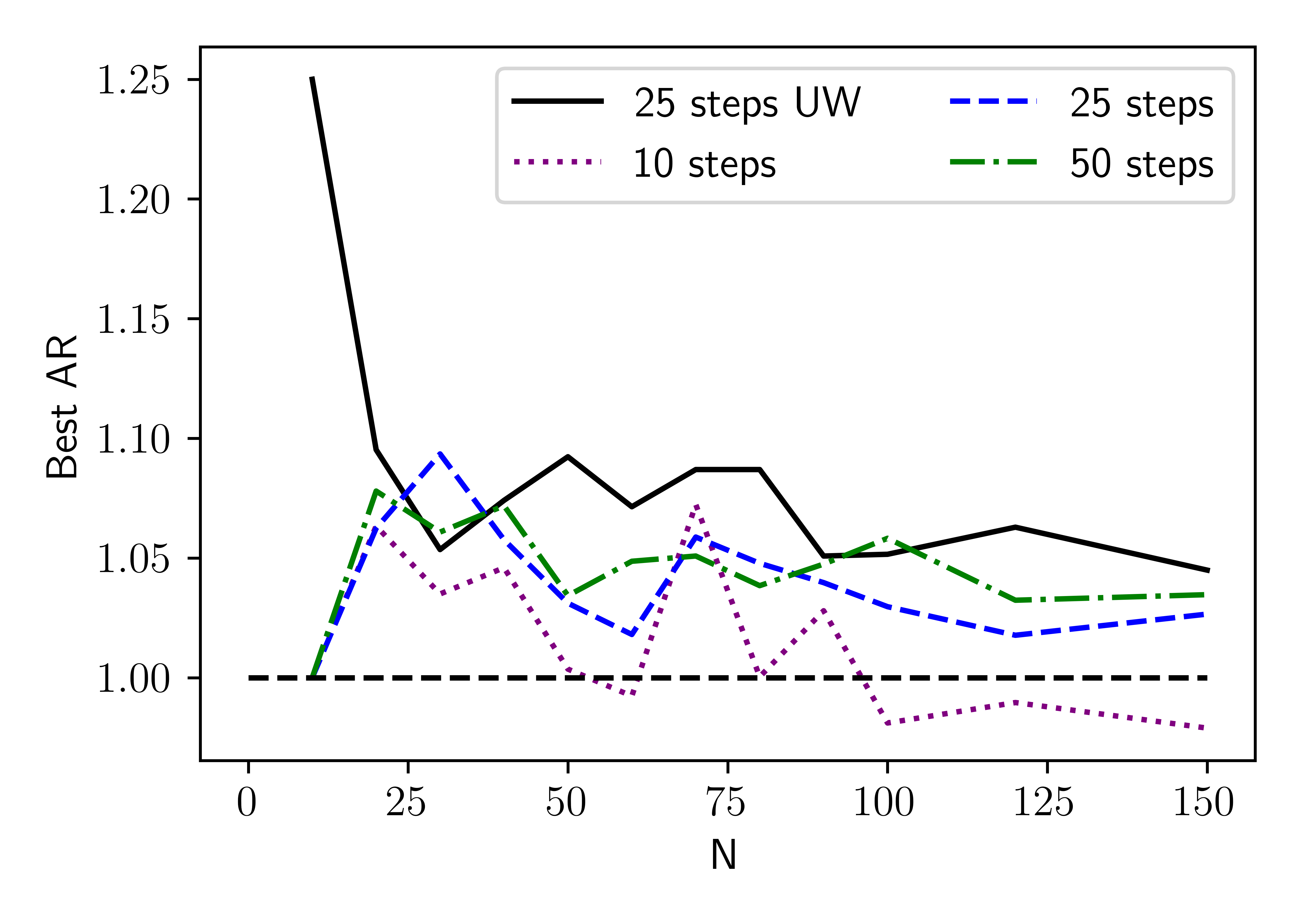} 
\caption{QITE results for the weighted MaxCut problem showing the best approximation ratio, computed by dividing the energy from QITE by the energy from GW, for $N$-qubit weighted graphs. Results are shown for 10 (dotted purple), 25 (dashed blue), and 50 QITE steps (dashed-dotted green), in addition to 25 steps for the unweighted (UW) case (solid black).  }   
    \label{fig:f1Bbest} 
\end{figure}  
\begin{figure}[ht!] 
    \centering
    \includegraphics[scale=0.48]{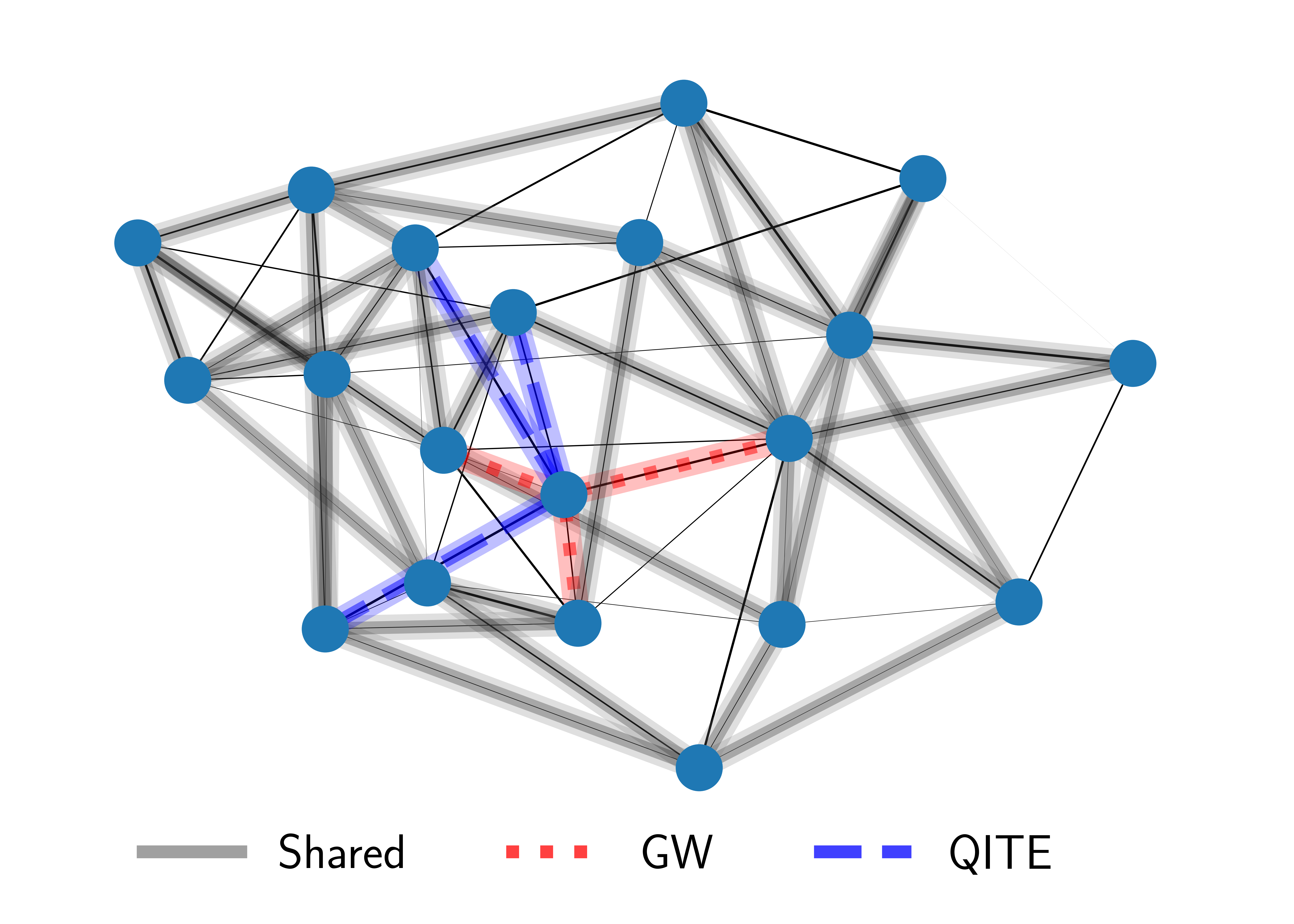} 
\caption{Example 20-vertex graph where 50-step QITE outperforms GW for the weighted MaxCut problem. The cuts common to GW and QITE are solid lines highlighted in grey. The cuts unique by QITE are dashed lines highlighted in blue, while the cuts unique to GW are  dotted lines highlighted in red.   }  
    \label{fig:f1C} 
\end{figure}  
First, we perform QITE on graphs with $6\leq N\leq 30$ vertices and measure the average number of random subgraphs to obtain convergence (P(GS)$>0.995$ overlap) to the ground state. The algorithm is performed repeatedly with different selections for the ITD edge and different random initial states until all attempted graphs have converged to the ground state. We compute this result for 25 and 50 QITE steps on weighted graphs, and also 25 QITE steps on unweighted graphs. We compute this as a function of the number of graph vertices $N$, shown in Fig.\ \ref{fig:f1A}. 

\begin{figure}[ht!] 
    \centering
      \includegraphics[scale=  0.5]{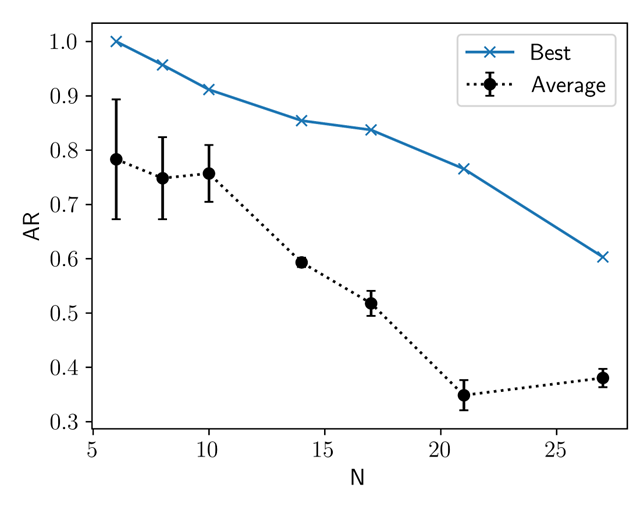} 
\caption{QITE results for the LABS problem showing the average approximation ratio (black circles) and best approximation ratio (blue $\times$) for 40 QITE steps using the linear Ansatz for problem size $6\leq N\leq 28$ for 100 different random initial states. The bars denote 1 standard error. } 
    \label{fig:f2A} 
\end{figure}  
\begin{figure}[ht!] 
    \centering
        \includegraphics[scale=  0.5]{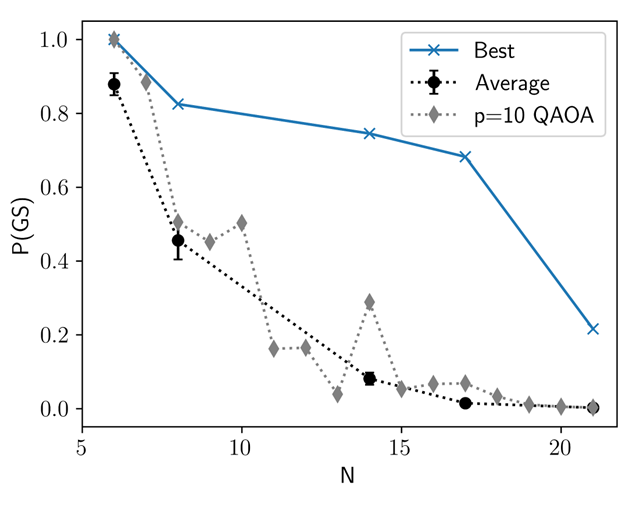}        
\caption{QITE results for the LABS problem showing the average (black circles) and best (blue $\times$) probability of measuring the ground state P(GS) for problem size $6\leq N\leq 18$ for 40 QITE steps using the linear Ansatz. The results from \cite{shaydulin2023evidence} for $p=10$ QAOA   are also shown (grey diamonds). } 
    \label{fig:f2B} 
\end{figure}  

We observe similar results for 25 and 50 steps on weighted graphs for $N<18$, while the unweighted graphs require less resources. For $N>18$, the results for 25 and 50 QITE steps begin to diverge, with almost twice the resources required at $N=30$. At this point, the energy gap between the ground state and first excited state is on the order of 0.01.   

For larger graphs with $N\geq  20$, the state produced from QITE sometimes converges to a state with lower energy than the lowest energy obtained from GW. To account for this, we compute the approximation ratio, the QITE energy divided by the GW energy, for graphs with $6<N<150$ vertices for single selections of random subgraphs and initial states, shown in Fig.\ \ref{fig:f1B}.

For 10 QITE steps, the average AR is less than $0.90$ for $N>10$, indicating that 10 steps are not sufficient to solve larger MaxCut problems. For 25 and 50 steps,  we observe the AR initially decrease with graph size to a minimum $\sim$0.90 for $N=50$ vertices, then increase for larger values of $N$ to a maximum value of $\sim$0.96 (0.94) for $N=125$ vertices at 50 (25) steps. We attribute the decrease to QITE performing better on smaller $N$ when GW produces the exact solution, and the increase to GW finding sub-optimal solutions when QITE produces a state with a finite overlap with a lower energy solution. To this end, we also plot the best AR for each number of steps in Fig.\ \ref{fig:f1Bbest}. For $N>10$, 25 and 50 vertex QITE remains above $1$, indicating that QITE produces solutions that are better than those found by the GW classical algorithm. 10-step QITE has a best AR $<1$ for $N>100$, indicating it produces solutions with higher energies than GW. For unweighted graphs, the average AR remains in the range of 0.92--0.95 for all $N$, and has a best AR $>1.05$ for $N>30$. 

An example of a 20-vertex weighted graph where 50-step  QITE outperforms the GW classical algorithm is shown in Fig.\ \ref{fig:f1C}. The cuts chosen by both algorithms are indicated, in addition to the cuts unique to QITE and GW. QITE chooses a solution with 3 different cuts than GW does. The difference in energy between QITE and GW is $\Delta E=-0.0404$. 


QITE with a linear Ansatz can be simulated efficiently classically, so the computation can be done for larger graph sizes than considered here without issue. Since here we chose 1 ITD edge, it would be interesting to see if choosing more ITD edges or choosing a different scheme would produce better results on larger graphs. We are currently investigating this in addition to the scaling of QITE on graphs with 500+ vertices. 
\section{LABS} \label{sec:4}

Here we discuss how we solve the Low Autocorrelation Binary Sequences (LABS) problem with QITE using both a linear and quadratic Ansatz, present results for problem size up to 28 qubits, and compare them with recent QAOA results \cite{shaydulin2023evidence}.

The goal of the LABS problem is to minimize the ``sidelobe  energy'' for a system of $N$ spins $\sigma_i\in\{+1,-1\}$ with autocorrelation $\mathcal{A}_k(\bm{\sigma})$ : 
\be \mathcal{E}_{\mathrm{sidelobe}}(\bm{\sigma})=\sum_{k=1}^{N-1}\mathcal{A}_k^2(\bm{\sigma}),~\mathcal{A}_k(\bm{\sigma})=\sum_{k=1}^{N-k} \sigma_i \sigma_{i+k}\ee 
This can be mapped to the ground state problem of the quantum Hamiltonian 
\bea  \bm{\mathcal{H}}^{\mathrm{LABS}} &=& 2\sum_{i=1}^{N-3}\sum_{t=1}^{\frac{N-i-1}{2}}\sum_{k=1}^{N-i-t}Z_iZ_{i+t}Z_{i+k}Z_{i+t+k} \nonumber\\
&& +\sum_{i=1}^{N-2}\sum_{k=1}^{\frac{N-i}{2}}Z_iZ_{i+2k}\eea
The quality of a solution can be quantified by the overlap with the ground states, $P(\bm{\sigma})$, and the ratio of the measured cost function over the the exact solution, $\braket{ \bm{\mathcal{H}} ^{\mathrm{LABS}}}/C_{\mathrm{max}}$. Note that in the literature, the merit factor is also used to quantify the quality of the solution $\ket{\psi}$, given by  
\be\mathcal{F}(\ket{\psi})=\frac{N^2}{2\langle \bm{\mathcal{H}} \rangle}=\sum_{\mathbf{s}}P(\bm{\sigma})\mathcal{F}(\bm{\sigma}),~\mathcal{F}(\bm{\sigma})=\frac{N^2}{2\mathcal{E}_{\mathrm{sidelobe}} (\bm{\sigma)}}\ee 
As with weighted MaxCut, we introduce imaginary time dependence to improve convergence to the ground state instead of an excited state. Since QITE with a linear Ansatz performs well with a Hamiltonian with quadratic terms, we introduce an ITD coefficient $\alpha[s]$ to the quartic terms of the Hamiltonian, and define the ITD Hamiltonian  
\bea  \bm{\mathcal{H}}^{\mathrm{LABS}} [s] &=& 2\alpha[s]\sum_{i=1}^{N-3}\sum_{t=1}^{\frac{N-i-1}{2}}\sum_{k=1}^{N-i-t}Z_iZ_{i+t}Z_{i+k}Z_{i+t+k} \nonumber\\
&&+\sum_{i=1}^{N-2}\sum_{k=1}^{\frac{N-i}{2}}Z_iZ_{i+2k}\ ,\ \ \alpha[s]= \frac{a  \lfloor  \frac{s}{a}  \rfloor}{n_{\mathrm{steps}}}   \eea 
where $s$ are the integer time indices ($s=0,\dots, n_{\mathrm{steps}}$), and $a$ determines whether $\alpha[s]$ is piecewise or linear ($a=1$) with respect to $s$. 

To improve the results further, we can add another time-dependent parameter, $\beta[s]$, that depends on the range of the quartic interaction terms, slowly adding increasingly nonlocal Hamiltonian parameters. 
\bea  \bm{\mathcal{H}}^{\mathrm{LABS}} [s] &=& 2\alpha[s]\sum_{i,t,k}\beta_{itk}[s]Z_iZ_{i+t}Z_{i+k}Z_{i+t+k} \nonumber\\ &&+\sum_{i,k} Z_iZ_{i+2k} \ , \label{eq:labsht}\eea 
where
\be  \beta_{itk}[s]=\begin{cases} \frac{b  \lfloor  \frac{s}{b}  \rfloor }{n_{\mathrm{steps}}}  & ,\,   \mathrm{max}(t,k,t+k,|t-k|)>R_{\mathrm{max}} \\ 1  &  , \, \mathrm{otherwise} \end{cases} \ee 
We reference a LABS solution bank \cite{OPUS2-git_labs-Boskovic} when computing the approximation ratio and the overlap with the ground state, P(GS), which is a quantum metric. Using Eq.\ \eqref{eq:labsht}, we perform the QITE algorithm with 40 steps on 50 random initial states comprised of $\ket{0}$ and $\ket{+}$ for $6\leq N\leq 21$. We first compute the approximation ratio by dividing the resulting QITE energy by the energy obtained from the solution bank.   The average and maximum AR for the 50 initial states are plotted in Fig.\ \ref{fig:f2A}.  The probability of measuring the ground state P(GS) is given in Fig.\ \ref{fig:f2B}. 
We plot the QITE results alongside the P(GS) values for $p=10$ QAOA obtained by \cite{shaydulin2023evidence}. We find the average P(GS) of the linear Ansatz QITE comparable to the $p=10$ QAOA, while the best case P(GS) consistently exceeds both values.   In Ref.\ \cite{shaydulin2023evidence}, LABS was solved exactly with simulated noiseless QAOA for problem size up to $N=40$. However, the authors used QAOA level up to $p=40$, which results in a gate depth on the order of $10^3$ for problem size $N=18$. This is challenging to execute on NISQ devices or classical simulators. For comparison, 
linear Ansatz QITE requires shallow quantum circuits and readily available classical computing resources. 
 
   Since the linear Ansatz QITE does not give P(GS)$=1$ as in weighted MaxCut, we considered a higher-order quadratic Ansatz. The details of the formulation are as follows.
   For the quadratic Ansatz, we used one containing the linear terms in addition to terms quadratic in $Y$: 
\be A_{\mathrm{quad}}[s] = \sum_{j\in V} a_{0,j}[s] Y_j + \sum_{i,j\in V} a_{i,j}[s] Y_i Y_j \label{eq:quadansatz}\ee 
It is convenient to define the vector consisting of all operators appearing in Eq.\ \eqref{eq:quadansatz}, $\bm{\mathcal{Y}} \equiv(Y_1,Y_2,\dots,Y_N,Y_1Y_2,Y_1Y_3,\dots,Y_1Y_N,Y_2Y_3,\dots)$, where the quadratic terms contain all unique choices of pairs of vertices. Evidently, this vector is of length $N(N+1)/2$. 
 
  \begin{figure}[ht!] 
    \centering
      \includegraphics[width=  0.48\textwidth]{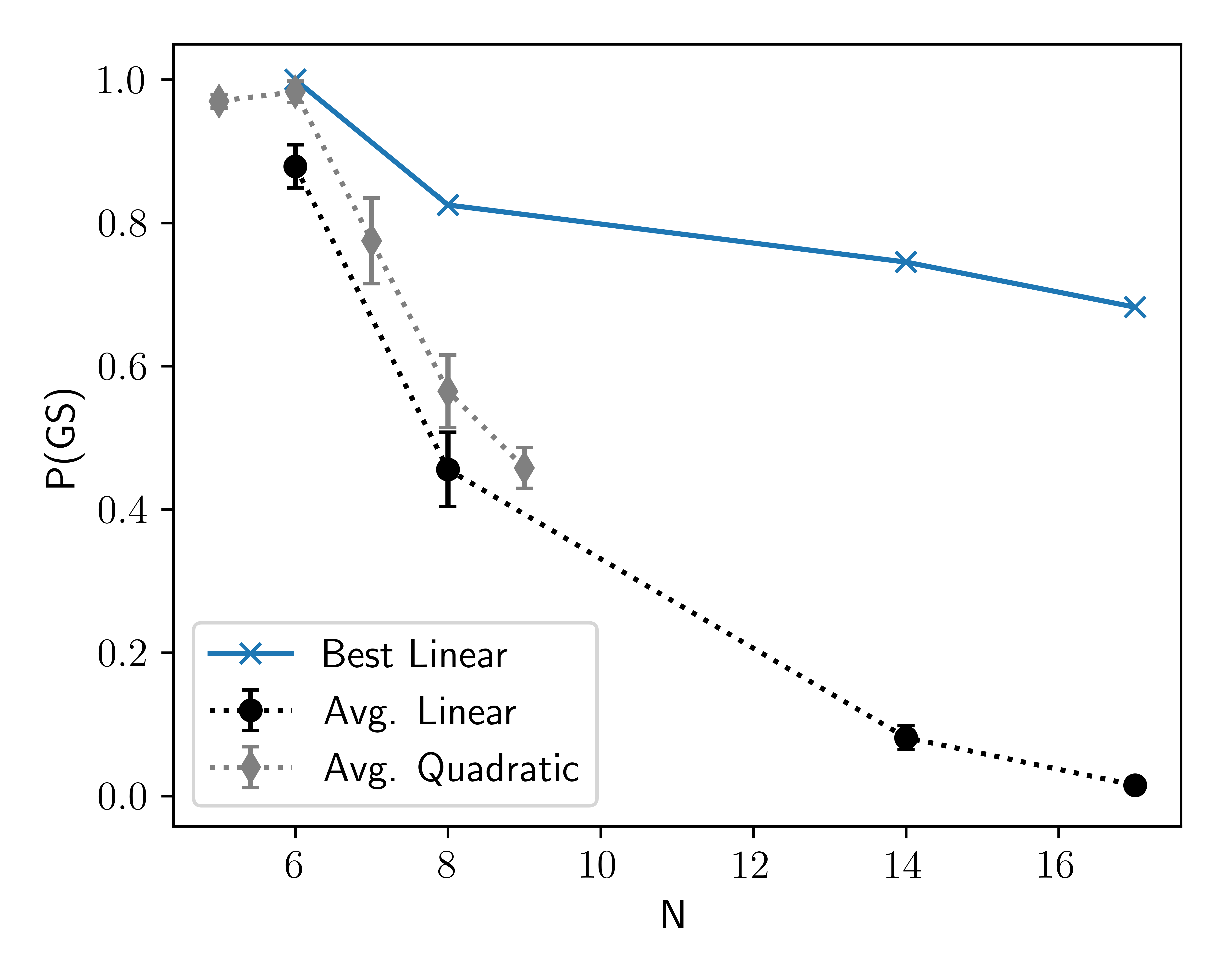} 
\caption{ Probability of measuring the ground state P(GS) after 40 QITE steps, shown for problem size $6\leq N\leq 18$ for the linear Ansatz, and $5\leq N\leq 9$ for the quadratic Ansatz. For the linear Ansatz,  the average P(GS) is shown in black circles, and the best blue  P(GS) is shown in blue $\times$. For the quadratic Ansatz, the average P(GS) is shown in grey diamonds. 
  } 
    \label{fig:f3} 
\end{figure}  

As in the linear Ansatz case, the coefficients $a_I[s]$, where $I = \{ i,j \}$, obey the linear system of equations
\be \bm{S}\cdot \bm{a} = \bm{b} \ee 
where we defined  \be 
S_{IJ}[s]=\braket{ \mathcal{Y}_I \mathcal{Y}_J } \ , \ \    b_J[s] = -\frac{i}{2} \braket{ [ \bm{\mathcal{H}}^{\mathrm{LABS}}  ,  \mathcal{Y}_J] } \ee
where all expectation values are evaluated with respect to the state $|\Psi[s-1]\rangle$ obtained in the previous step. If the initial state is chosen to be the tensor product of eigenstates of $X$ and $Z$, then we have $\bm{S}=\mathbb{I}$, and therefore $\bm{a}=\bm{b}$. 
The algorithm then proceeds as in the linear Ansatz case with updates of the form 
\be \ket{\Psi [s]} = e^{-i\tau A_{\mathrm{quad}}[s]} \ket{\Psi[s-1]} \label{eq:approxa}\ee
The quadratic Ansatz gives a minor improvement in the P(GS) over the linear Ansatz,  as shown in Fig.\ \ref{fig:f3}. Therefore, we do not see a benefit to including higher-order terms in the Ansatz in the range of $5<N<10$. It is conceivable that there may be benefits at larger problems sizes beyond our current resources to test.   

In \cite{shaydulin2023evidence}, the scaling advantage of QAOA over classical solvers was found in the range of $28\leq  N\leq 40$, which is larger than the problem sizes we tested. Therefore, we need to investigate the performance of linear and quadratic Ansatz QITE compared to QAOA and classical algorithms in this range. 
The quadratic Ansatz simulation requires significantly more computational resources and precision than the linear Ansatz, but we are currently working on extending our results to larger problem sizes.

\section{Conclusion}\label{sec:V} 
The QITE algorithm effectively cools a system to zero temperature at which the system settles to its ground state reaching the lowest energy level of its Hamiltonian. By encoding combinatorial optimization problems in terms of a quantum Hamiltonian, one can solve these problems by finding the ground state of the Hamiltonian corresponding to the optimal solution. QITE allows for a flexible Ansatz which can be chosen to be separable or entangling. Although solutions of  combinatorial optimization problems involve separable states, quantum algorithms, such as QAOA, introduce entangling operations to solve these problems. 

In this work, we investigated the performance of QITE on PUBO problems, concentrating on the weighted MaxCut and LABS combinatorial optimization problems. Our method was a generalization of the approach introduced in Ref.\ \cite{Alam2023} that was successfully applied to unweighted MaxCut. In general, we expected an increased difficulty in PUBO cases due to smaller gaps between the ground and first excited state energies. We compared the performance of QITE on NWS graphs with up to 150 vertices with the commonly used classical algorithm GW, which is widely believed to offer the best performance guarantee. We found that QITE with a linear Ansatz and ITD edges often outperforms the classical GW algorithm on weighted MaxCut instances after a sufficient number of steps, yielding an average AR  $\sim0.95$ for $N>100$ vertices.  Our separable Ansatz can be simulated efficiently classically. In Ref.\ \cite{crooks2018performance}, the performance of noiseless QAOA was assessed on MaxCut problems up to 17 vertices, and was observed that level $p=5$ was required for results to be comparable with GW. Especially with larger graphs, this is outside of the range of NISQ devices. In Ref.\ \cite{Guerreschi_2019}, it was estimated that graph sizes of several hundreds to thousands of vertices are required for quantum advantage over classical solvers on the MaxCut problem.  Our results indicate that much larger graphs are needed to be considered in order to observe quantum advantage. Given the attendant increase in the depth of the quantum circuit, a quantum computation may not be suitable on NISQ devices. It is important to analyze the applicability of our method further to better assess the utility of NISQ devices.

We also analyzed the performance of QITE on the LABS problem. This is a PUBO probem, as the quantum Hamiltonian encoding the LABS problem includes quartic terms in the Pauli $Z$ matrix. We tested the performance of QITE with a separable linear Ansatz and a randomly chosen separable initial state.  We compared to known solutions of the problem from a solution bank, and calculated the AR and probability of measuring the ground state, P(GS). Although in general we did not obtain convergence to the ground state, we found that the average P(GS) of the linear Ansatz QITE with 40 steps was comparable to $p=10$ QAOA \cite{shaydulin2023evidence}. More importantly, the best case P(GS) for each problem size was consistently higher than the QAOA results, indicating that an appropriate choice of initial state gives a high probability of solving the LABS problem. 

Expecting improvement and possibly quantum advantage to be found using an entangling Ansatz containing higher-order terms, we considered an entangling Ansatz with both linear and quadratic terms. Comparing it to the separable linear Ansatz, we found that the entangling Ansatz gave a P(GS) which was no more than 10\% higher than the result from the linear Ansatz for problem sizes up to $N=9$. Thus, the entangling quadratic Ansatz did not perform significantly better than the separable linear Ansatz in the problem size range we tested, indicating that we might need to explore larger problem sizes to see a significant advantage due to quantum entanglement. 

Obtaining quantum advantage in combinatorial optimization problems appears to require larger problem sizes than those that can be handled by NISQ devices, or, perhaps, more complicated PUBO problems than the ones studied here. Both research directions, i.e., searching for a combinatorial optimization problem where an entangling Ansatz outperforms the linear Ansatz significantly even at sufficiently small problem sizes implementable on NISQ devices, as well as simulating large enough problem among those considered here where an entangling Ansatz outperforms the linear Ansatz, are currently being pursued. 

 
  \acknowledgments
  We thank Phillip C.\ Lotshaw for useful discussions. This work was supported by the DARPA ONISQ program under award W911NF-20-2-0051, and NSF award DGE-2152168.


%

\end{document}